\documentclass[runningheads]{llncs}
\usepackage{graphicx}
\usepackage{multirow}
\usepackage{todonotes}
\usepackage{xcolor}
\usepackage{comment}
\usepackage{adjustbox}
\usepackage{amsmath}
\usepackage[utf8]{inputenc}
\usepackage[english]{babel}
\usepackage{color}
\usepackage{amsfonts}
\usepackage{url}
\usepackage[mathscr]{euscript}
\let\euscr\mathscr \let\mathscr\relax% just so we can load this and rsfs
\usepackage[scr]{rsfso}
\usepackage{caption}
\usepackage[labelfont=bf]{caption}
\usepackage{tabularx,booktabs}
\newcolumntype{C}{>{\centering\arraybackslash}X} % centered version of "X" type
\usepackage{multicol}
\usepackage{algorithm2e}
\usepackage{booktabs,ragged2e}
\usepackage[utf8]{inputenc}
\usepackage{todonotes}
\usepackage{lineno}
\usepackage{makecell}

\usepackage[flushleft]{threeparttable}
\usepackage{subcaption}
\usepackage{color}

\begin{document}
%-------------------
\title{EndoUDA: A modality independent segmentation approach for endoscopy imaging}
\titlerunning{EndoUDA: Modality independent segmentation}
\author{
Numan Celik \inst{1,2}\orcidID{0000-0003-1813-1036}, Sharib Ali \inst{1,2,3}\orcidID{0000-0003-1313-3542}, Soumya Gupta \inst{1,2}\orcidID{0000-0001-5874-5273}, Barbara Braden \inst{2,3}\orcidID{0000-0002-8534-6873}, Jens Rittscher \inst{1,2,3}\orcidID{0000-0002-8528-8298}}
\authorrunning{N. Celik, S. Ali et al.}
\institute{Department of Engineering Science, Institute of Biomedical Engineering, University of Oxford, Oxford, UK \and Big Data Institute, University of Oxford, Li Ka Shing Centre for Health Information and Discovery, Oxford, UK
\and NIHR Oxford Biomedical Research Centre, Oxford, UK \and Translational Gastroenterology Unit, Experimental Medicine Div., John Radcliffe Hospital, University of Oxford, Oxford, UK\\
\email{\{numan.celik, sharib.ali, jens.rittscher\}@eng.ox.ac.uk}}
% Author list: Numan Celik, Sharib Ali, Soumya Gupta, Barbara Braden, Jens Rittscher
\maketitle      
%-------------------
\begin{abstract}
%------------------
% Overview of motivation of the study 
Gastrointestinal (GI) cancer precursors require frequent monitoring for risk stratification of patients. Automated segmentation methods can help to assess risk areas more accurately, and assist in therapeutic procedures or even removal. In clinical practice, addition to the conventional white-light imaging (WLI), complimentary modalities such as narrow-band imaging (NBI) and fluorescence imaging are used. While, today most segmentation approaches are supervised and only concentrated on a single modality dataset, this work exploits to use a target-independent unsupervised domain adaptation (UDA) technique that is capable to generalize to an unseen target modality. In this context, we propose a novel UDA-based segmentation method that couples the variational autoencoder and U-Net with a common EfficientNet-B4 backbone, and uses a joint loss for latent-space optimization for target samples. We show that our model can generalize to unseen target NBI (target) modality when trained using only WLI (source) modality. Our experiments on both upper and lower GI endoscopy data show the effectiveness of our approach compared to naive supervised approach and state-of-the-art UDA segmentation methods.

\end{abstract}
\section{Introduction}

Most gastrointestinal (GI) cancers are preventable. In 2018, the five major types of GI cancers, which include those of the stomach, liver, oesophagus, and colorectum account for 26\% of the global cancer incidence and 35\% of all cancer deaths~\cite{arnold2020global}. Endoscopy, a vital tool for screening and disease surveillance, is however heavily operator dependent and 12\% of cancers are missed~\cite{menon2014commonly}. Artificial intelligence-based systems can play a vital role in improving diagnostic quality. However, it is critical that these tools integrate directly into the existing clinical workflow and generalize well to the previously unseen data. 

Today, high-definition endoscopes provide a sufficient resolution to allow for a detailed visualisation of the mucosal surface. In addition to the conventional white light imaging (WLI) modality, autofluorescence imaging, and electronic chemoendoscopy techniques such as narrow-band imaging have been developed to improve the early detection of cancer and its precursors~\cite{subramanian2014advanced}. It is important that any computer assisted methods should work seamlessly in these different modalities and avoid the need for any modality-specific training. Developing such a modality agnostic approach for detection and segmentation of pre-cancerous changes in the esophagus, and polyps in the lower GI is the focus of this paper.  

\indent{Convolutional} Neural Network (CNN) based approaches~\cite{FCN,Olaf:MICCAI2015,10.1007/978-3-030-01234-2_49}, all  well established in biomedical imaging, have been applied to segmentation tasks in GI endoscopy ~\cite{ALI2021102002,guo2020polyp,wu2021elnet}. Wu et al.~\cite{wu2021elnet} proposed a dual neural network based on U-Net with ResNet backbone for automatic esophageal lesion segmentation. Guo et al.~\cite{guo2020polyp} utilized {a} FCN-based architecture with atrous kernels to segment polyps. However, all these supervised approaches are trained on WLI endoscopy and do not support working in different imaging modalities. 
\begin{figure}[t!]
\centering
\includegraphics[width=0.99\textwidth]{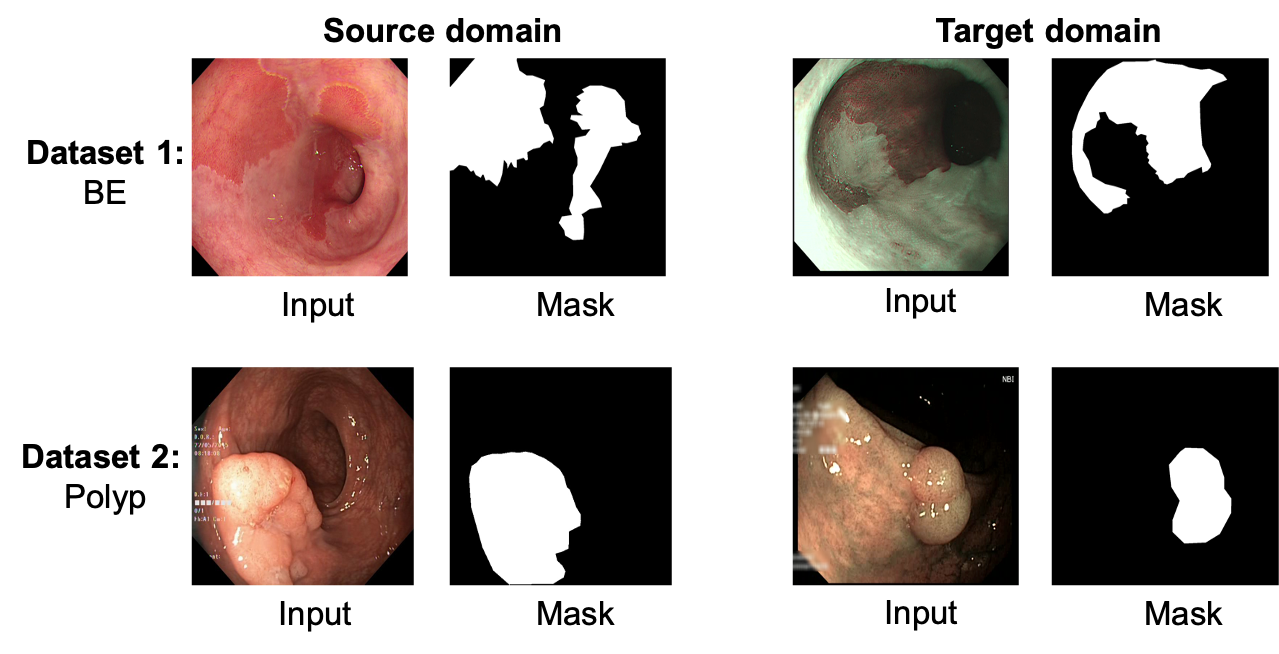}
\caption{Example images and masks of the two GI datasets for which source and target domains are separated based on their modality.~\label{fig:1}}
% The top row illustrates the Barrett's esophagus while the bottom row represents the polyps. Here, the left column correspond to white light imaging (WLI) modality which is source domain in our study and the right column demonstrate the narrow band imaging (NBI) modality used as target domain.
\end{figure}
To date, the problem of quantifying disease related changes in NBI, which can produce a very different appearance of the mucosal surface when compared to WLI (see Fig.~\ref{fig:1}) has not been addressed. Training methods for different imaging modalities independently is prohibitive. Firstly, extensive training datasets that cover a similar range of conditions would need to be curated to ensure a comparable performance. Secondly, manual annotations by human experts are required which is not only labour intensive but also expensive. 
To tackle these issues domain adaptation methods have been used to adapt trained models to unseen target data for segmentation~\cite{Tsai,CardioUDA,pmlr-v121-haq20a,10.1007/978-3-030-58539-6_25}. 
{Although these methods promise to generalise, so far there are only a few examples of applying such methods to medical data ~\cite{CardioUDA,pmlr-v121-haq20a}}. Additionally, the available methods still under-perform and have not been adapted for extremely diverse modality cases such as that in the GI endoscopy. 

%Tsai et al.~\cite{Tsai} developed a domain adaptation model for semantic segmentation based on a multi-level adversarial network that shares the segmentation loss for both source and target domains. In the discriminator network, target domain predictions are matched to the source domain predictions whether the input is from the source or target domain via an adversarial loss. In the context of biomedical image segmentation, Dong et al.~\cite{CardioUDA} proposed an unsupervised domain adaptation (UDA) technique for cardiothoracic ratio estimation by performing domain invariant chest organ segmentation. Haq et al.~\cite{pmlr-v121-haq20a} introduced an alternative UDA framework that included a decoder network to reconstruct the input images for cell segmentation. Pandey et al.~\cite{10.1007/978-3-030-58539-6_25} employed UDA approach to segment human skin under different imaging modalities via a variational auto-encoder (VAE) based generative model.
%

To develop a modality-agnostic approach, we propose to use an unsupervised domain adaptation (UDA) technique that can be trained on available single WLI modality dataset and can be applied on an unseen NBI imagery. For such settings, where classical supervised models fail to generalize and perform poorly, we show that our proposed target-independent UDA technique, EndoUDA, has the potential to generalize and {to} provide improved segmentation on NBI modality for both BE and polyp datasets. 
%
%Mendel et al.~\cite{mendel2017barrett} used transfer learning based method to learn features of BE in early detection of EAC. 
%\section{Related work} % Or just embed in introduction
%%%%% done till here Time stamp:12:22, Tuesday
%%%%% it was Fig_2.png
\begin{figure}[t!]
\centering
\includegraphics[width=0.99\textwidth]{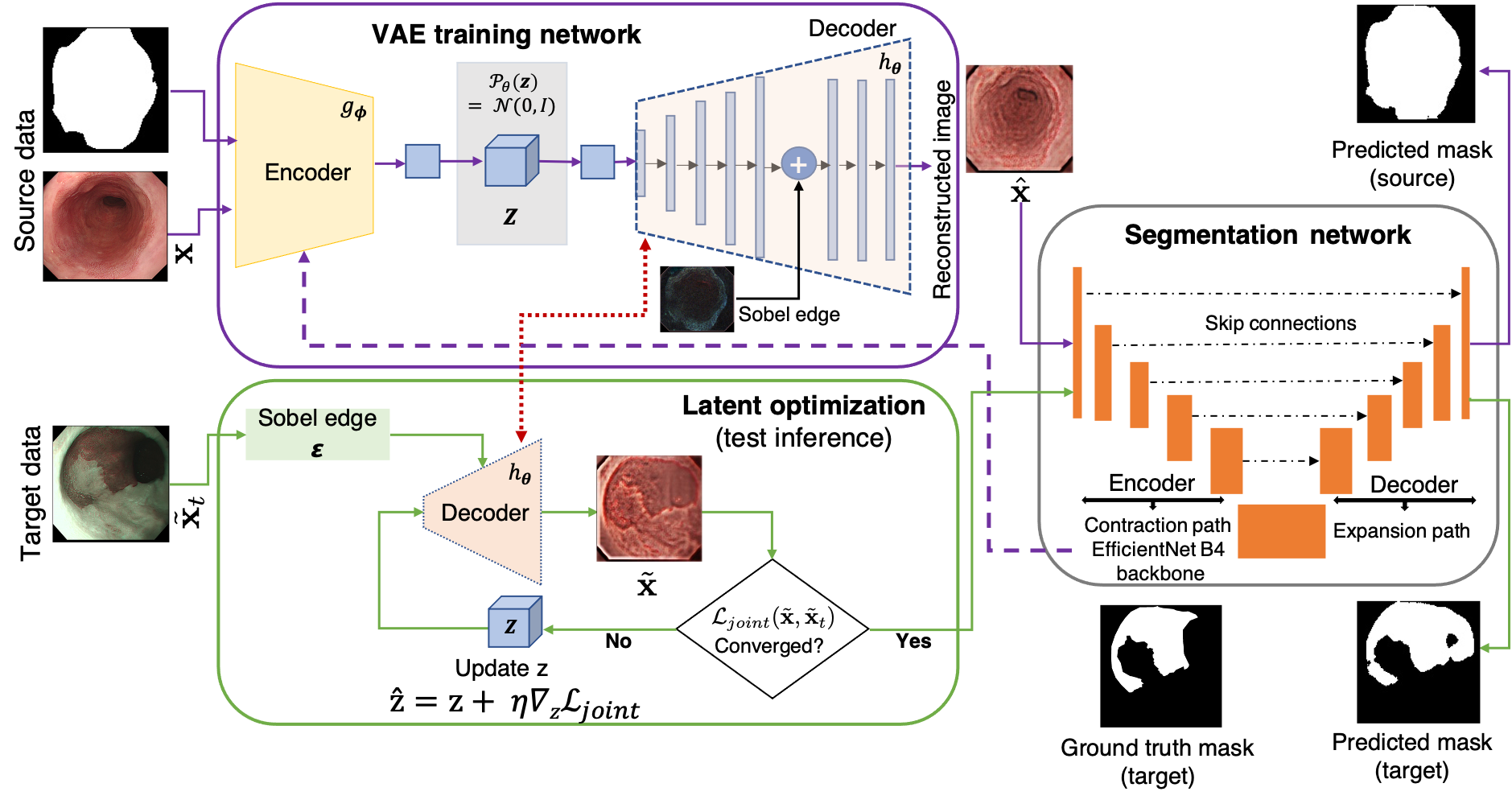}
\caption{Complete  architecture  of  the proposed EndoUDA: a target-independent unsupervised  domain  adaptation.~\label{fig:2}}
\end{figure}
%%%%%
\section{Method}{\label{section:method}}
%%%%%
% Here, the method should be straight. You already have contributions above. Readers want to see only details in the method. Rewritting!

Our proposed EndoUDA consists of components for training and test time inference shown in Figure~\ref{fig:2}. 
The training network, utilizes a shared EfficientNet-B4 encoder for our variational autoencoder (VAE) and a U-Net-based segmentation model. The VAE module allows to learn the source domain representations while U-Net module leverages the learned encoded features for semantic segmentation. For inference at test time, learnt latent space embeddings are further optimized to translate to the unseen modality dataset by continuously updating via a joint loss minimization scheme. Below, details of each of these modules are presented. 
%
%This study has two main contributions in the novelty of the proposed methodology: first improving the network architecture by coupling encoder of the segmentation network into the encoder of the VAE model during the training; second introducing a normalized cross-correlation (NCC) loss in the latent search algorithm during inference to the target domain. The proposed EndoUDA approach is based on the UDA framework which consists of three main modules: (i) adversarial training module on source domain utilizing VAE generative model (Fig.~\ref{fig:2b}),ii) segmentation module based on Efficient-Net B4 backbone U-Net architecture and (iii) inference module on target domain using VAE latent search algorithm by optimising a mixture loss of normalized cross correlation \(L_{NCC}\) and structural similarity score\(L_{ssim}\) (Fig.~\ref{fig:2b}). 
% TODO: ADD (also in abstract!)
% {\color{red}We use a target-independent approach which means that the network is only trained on the source domain.} By improving the UDA architecture~\cite{pandey2020unsupervised}, we propose a semantic segmentation framework based on generative model to learn invariant features from source domain which is built by WLI only, to provide automatic segmentation of Barrett's area and neoplastic lesions for unlabelled target dataset which is deployed by NBI.

\subsection{VAE training on source domain}
VAE is a latent space derived generative model that is based on variational inference distributions. Let $\euscr{P}_\theta (\textbf{z}\vert \textbf{x})$ be the latent space distribution for sample $\textbf{x}$ and $\mathcal{Q}_\phi (\textbf{z}\vert \textbf{x})$ be an identical tractable distribution so that it approximates the true posterior \begin{math}\euscr{P}_\theta (\textbf{z}\vert \textbf{x})\end{math}. Then, the variational lower bound term of the loss $\euscr{L}(\theta,\phi)$ between probabilistic $h_\phi$ (encoder, outputs the parameters $\mu_\textbf{z}$ and $\sigma_\textbf{z}$ of a distribution) and $h_\theta$ (decoder) neural networks. In Eq.~(\ref{eq:2}), approximated $\mathcal{Q}_\phi (\textbf{z}\vert \textbf{x})$  eliminates the non-negative KL-divergence term {imposing} a lower bound on the data log-likelihood only. This is maximized {by the} VAE {via} parameterizing $\euscr{Q}_\phi(\textbf{z}\vert \textbf{x})$ and  $\euscr{P}_\theta (\textbf{x}\vert \textbf{z})$ which is similar to the minimization of the reconstruction loss $\euscr{L}_r(\textbf{x})$ in Eq.~(\ref{eq:1reconstruction}).  
\begin{comment}
\begin{equation}{\label{eq:2}}
\ln{(\euscr{P}_\theta (\textbf{x}))} = \euscr{L}(\theta,\phi) + \mathbb{D}_\textit{KL}[\euscr{Q}_\phi(\textbf{z}\vert \textbf{x})\|\euscr{P}_\theta (\textbf{z})]
\end{equation}
\end{comment}
%where,
\begin{equation}{\label{eq:2}}
\euscr{L}(\theta,\phi) = \mathbb{E}_\euscr{Q_\phi}[\ln{(\euscr{P}_\theta (\textbf{x}\vert \textbf{z}))}] - \mathbb{D}_\textit{KL}[\euscr{Q}_{\phi}(\textbf{z}\vert \textbf{x})\|\euscr{P}_{\theta} (\textbf{z}\vert \textbf{x})] %\euscr{P}_\theta (x\vert z)
\end{equation}
%where KL$(\cdot\|\cdot)$ defines KL-divergence; $\euscr{L}(\theta,\phi)$ denotes the lower bound on the image likelihood 

Compared to the VAE framework in~\cite{10.1007/978-3-030-58539-6_25}, we propose to couple the encoder layer of VAE as the segmentation network backbone. Based on our experiments, we have chosen {an} EfficientNet which computes the approximate posterior $\mathcal{Q}_\phi (\textbf{z}\vert \textbf{x})$. Unlike {Pandey {\it et al.}} ~\cite{10.1007/978-3-030-58539-6_25}, we {remove} the originally used perceptual loss module in the decoder layer used to compute the image likelihood of the observed data \begin{math}\euscr{P}_\theta (\textbf{x}\vert \textbf{z})\end{math}. The decoder network of {the} VAE contains 5 convolution layers followed by a LeakyRelu function (Leak=0.2) and {a \textit{tanh} activation function for the reconstructed output in the final convolution layer}.
{A} Sobel filter generated edge map is concatenated with the sixth layer of the decoder network of {the} VAE through a \textit{tanh} nonlinear function (see Fig.~\ref{fig:2}). The reconstruction loss $\euscr{L}_r(\textbf{x})$ is minimized for the VAE training network given by:
\begin{equation}{\label{eq:1reconstruction}}
\euscr{L}_r({\textbf{x}}) = \euscr{L}(\theta,\phi) = \frac{1}{N}\sum_{i=1}^{N}\|\textbf{x}^i - h_\theta(h_\phi(\textbf{x}^i))\|^2,
\end{equation}
%
%Compared to the VAE framework presented in~\cite{pandey2020unsupervised}, we use propose to use two neural networks: as a new impact of this study, for the encoder network of VAE we couple the encoder of segmentation network (EfficientUNet) to compute the approximate posterior; and a decoder network to compute the image likelihood of the observed data \begin{math}\euscr{P}_\theta (\textbf{x}\vert \textbf{z})\end{math}. Differently from their study we eliminate perceptual network to simplify the pipeline. %During training, the ``reconstructed'' \textbf{x} image is sampled in a two-step process: (i) sample \textbf{z} $\sim$ $\euscr{N}$(0,~\textit{I}) and (ii) sample \textbf{x} from \begin{math}\euscr{P}_\theta (\textbf{x}\vert \textbf{z})\end{math} at the outer layer of the decoder network. 
% As shown in Eq. 2, the reconstruction loss (mean squared error) $\euscr{L}_r$ is also computed as an $l2-$norm between the source sample and the reconstructed sample.
%
\noindent where $\textbf{x}$ is the source domain image, $N$ is total number of samples,  $h_\phi$ and $h_\theta$ are encoder and decoder networks, respectively. Here, the reconstructed image is given by $\hat{\textbf{x}} = h_\theta(h_\phi(\textbf{x}))$.
In addition, as shown in Fig.~\ref{fig:2}, the edges of the input images are extracted and concatenated into the decoder network in VAE through a skip connection to better generalize the source and target domains. Fig.~\ref{fig:2} shows a schematic diagram of the complete VAE architecture used for training on the source domain. 
\subsection{Segmentation module}
The segmentation network consists of a U-Net with an EfficientNet-B4~\cite{pmlr-v97-tan19a} backbone. The input from the reconstructed image obtained from the decoder of the VAE is fed to the already tuned weights of EfficientNet-B4. The segmentation network is adapted by a supervised loss on the source domain given by a classically used binary cross-entropy loss ($\euscr{L}_{BCE}$) and dice coefficient loss  ($\euscr{L}_{DSC}$)~\cite{7785132}. $\euscr{L}_{DSC}$ is represented as {$\euscr{L}_{DSC} = (2\sum \textbf{p}\textbf{g})/(\sum \textbf{p}^2 + \sum \textbf{g}^2)$}, where $\textbf{p}$ is the predicted probability map and $\textbf{g}$ is the ground truth mask.
%The segmentation framework is shown in Fig.~\ref{fig:2b}.
%
% \begin{equation}{\label{eq:1}}
% \euscr{L}_{DSC} = \frac{2\sum pg}{\sum p^2 + \sum g^2},
% \end{equation}
% \noindent where $p$ is the predicted probability map and $g$ is the ground truth mask. 
%
\subsection{VAE latent search optimization via new joint Loss}
The trained VAE encoding aims to find the nearest point from the source domain, given a sample from the target domain, through an optimization process and updates of the latent space representation \textbf{z}, as can be seen in Fig.~\ref{fig:2}. Once the decoder $h_\theta$ of the VAE is trained on the source distribution $\euscr{P}_s(\textbf{x})$, given an image $\tilde{\textbf{x}}_t$ from the target distribution, the latent search algorithm finds the nearest point by optimizing latent vector over the iterative process. As shown in the Fig.~\ref{fig:1}, the WLI images are used in the source domain while the NBI are used in the target domain. For faster optimization convergence of the latent space encoding with improved generalization ability of the network, we propose to use a joint loss function that comprises of the normalized cross correlation (NCC) \(\euscr{L}_{NCC}\) and the structural similarity index measure (SSIM)~\cite{1284395} loss \(\euscr{L}_{ssim}\) between the image $\tilde{\textbf{x}}_t$ from target domain and the reconstructed target image $\tilde{\textbf{x}}$ from $h_\theta$. The loss function for computing NCC is defined as:
%%%%%%%%%
\begin{equation}{\label{eq:ncc}}
\euscr{L}_{NCC}(\tilde{\textbf{x}},\tilde{\textbf{x}}_t) = \frac{1}{2N}\sum(\frac{\tilde{\textbf{x}}-\mu_s}{\sqrt{\sigma_s^2+\epsilon^2}} - \frac{\tilde{\textbf{x}}_t-\mu_t}{\sqrt{\sigma_t^2+\epsilon^2}})^2,
\end{equation}
%%%%%%%%%
\noindent where $\tilde{\textbf{x}}$, $\mu_s$ and $\sigma_s$ are reconstructed images from decoder network, mean and standard deviation of the latent encoded image, respectively. Similarly, $\tilde{\textbf{x}}_t$, $\mu_t$ and $\sigma_t$ correspond to target image input, mean and standard deviation of the target image. The SSIM for a pair of images ($\tilde{\textbf{x}}$, $\tilde{\textbf{x}}_t$) and loss function $\euscr{L}_{ssim}$ is defined as:
%%%%%%%%%
\begin{equation}{\label{eq:ssim}}
%\begin{align}
\begin{split}
SSIM(\tilde{\textbf{x}}, \tilde{\textbf{x}}_t) &= l(\tilde{\textbf{x}}, \tilde{\textbf{x}}_t)^\alpha * c(\tilde{\textbf{x}}, \tilde{\textbf{x}}_t)^\beta * s(\tilde{\textbf{x}}, \tilde{\textbf{x}}_t)^\gamma,
\\
\euscr{L}_{ssim}(\tilde{\textbf{x}}, \tilde{\textbf{x}}_t) &= 1 - SSIM(\tilde{\textbf{x}}, \tilde{\textbf{x}}_t),
\end{split}
%\end{align}
\end{equation}
%%%%%%%%%
\noindent here \textit{l}, \textit{c} and \textit{s} denote luminance, contrast and structure similarities of the given pair of images, respectively with the parameters $\alpha$, $\beta$ and $\gamma$ $>$ 0. Also, the similarity loss function \(\euscr{L}_{ssim}\) is presented. The final joint loss function is given by:
%%%%%%%%%
% \begin{equation}
% \euscr{L}_{ssim}(\textbf{x},\hat{\textbf{x}}) = 1 - SSIM(\textbf{x},\hat{\textbf{x}})
% \end{equation}
%%%%%%%%%
\begin{equation}{\label{eq:joint}}
\euscr{L}_{joint} = \lambda \euscr{L}_{NCC} + (1-\lambda) \euscr{L}_{ssim},
\end{equation}
%%%%%%%%%
\noindent where $\lambda = 0.75$ is set empirically. After the convergence of \(\euscr{L}_{joint}\), the optimal latent space representation $\hat{\textbf{z}} = \textbf{z} + \eta \euscr{L}_{joint}$ is used to generate the closest clone of source domain with a learning rate $\eta$ which is then used as an input to the segmentation network to predict the target mask.
% \begin{equation}{\label{eq:zupdate}}
% \hat{\textbf{z}} = \textbf{z} + \eta \euscr{L}_{joint}
% \end{equation}
%
\section{Experiments and results}
\subsubsection{Implementation Details.}
%In our experiments, we use EfficientNet-B4 as the backbone for the UNet segmentation network and encoder of the VAE framework. The UNet segmentation network is trained on the source domain only. For the extraction edge information of the input image, we use Sobel operator in which then the extracted edge image is concatenated with the 6th layer of the decoder network of VAE through a \textit{tanh} nonlinear function as shown in Fig.~\ref{fig:2}. The VAE framework has two networks: an encoder and a decoder network. We couple EfficientNet backbone encoder of UNet network to the encoder of the VAE. The decoder network of VAE contains 5 convolution layers followed by a LeakyRelu function (Leak=0.2) and a final convolution layer with \textit{tanh} activation function for the reconstructed output. 

For the training of our VAE network we used RMSProp with learning rate (lr) of 0.0001 while for the segmentation module Adam optimizer was used with $\beta_1$=0.5, $\beta_2$=0.999, and $lr=0.001$. The initial learning rate was reduced by a factor of 10 for every validation loss that did not improve in the last 3 epochs and a stopping criteria was set for no improvement in validation loss upto 10 epochs. We used 100 epochs for training our EndoUDA training module with batch size of 64. For the test module of EndoUDA, we set learning rate of $\eta = 0.001$ for optimization of the latent space representation and a fixed 40 iterations with batch size of 32 was set. The proposed method was implemented using {the} Tensorflow framework on a machine deployed with NVIDIA RTX 2080Ti graphics card with 12 GB GPU memory. All input images were resized to 128 $\times$ 128 {pixels}.

%Adam optimizer is used with $\beta_1$=0.5, $\beta_2$=0.999, and learning rate, $lr=0.001$. and RMSProp (lr=0.0001) are used as optimization for the segmentation and VAE networks, respectively. The initial learning rate ($1.10^{-3}$) was reduced by a factor of 10 whenever the moving average of the validation loss has not improved in the last 3 epochs and training was stopped after no improvements in the last 10 epochs. VAE is trained using 100 epochs with batch size 64. The proposed method was implemented using Python in Tensorflow framework on a machine deployed with GTX2080Ti graphics card with 12 GB GPU memory. All input images were resized to 128 × 128.
\subsubsection{Datasets.}
We have used two gastrointestinal (GI) endoscopy datasets for analyzing two distinct endoscopically  found precancerous anomalies, namely, Barrett's esophagus (BE) and {polyps}. To evaluate the efficacy of the {proposed} EndoUDA method, we have used clinically acquired white light imaging (WLI) and narrow-band imaging (NBI) {modality} data for BE and polyps. BE dataset consists of 1182 endoscopy images acquired from 68 unique patients {of which} 611 WLI images {are} used as source domain data (train set: 90\%, validation set: 10\%) and 571 NBI images as target domain data {are} used for testing. Similarly, for {the} polyp dataset, we used {the} publicly available Hyper-Kvasir dataset~\cite{borgli2020} for training which consists of 1000 images (train set: 80\%, validation set: 20\%). {In addition, we} used clinically acquired 42 NBI images of polyps acquired from 20 patients at our local hospital and from collaborators.
%and to evaluate the effectiveness of the proposed EndoUDA pipeline. The first dataset is a new dataset and curated by us in collaboration with the John Radcliffe Hospital. This new dataset obtained from 68 patients contains 1182 endoscopy images in total, and dedicated on segmenting BE region for both source (WLI) and target domain (NBI). The source dataset contains 611 WLI sets of which 420 images were used for training and 191 images for validation. The target NBI dataset of BE consists of 571 images with the corresponding annotations. The annotations of the BE dataset were performed by two experienced post-doctoral researchers and reviewed by two senior gastroenterologists using VGG image annotator tool \cite{dutta2019vgg}. The source (WLI) of polyp dataset is obtained from Hyper-Kvasir dataset \cite{borgli2020} which is publicly available and contains 1000 images on which 800 images were used for training and 200 images for validation. The target dataset (NBI) of polyp contains 42 images. 
\label{sec:experiments}
%%%%%%%%%
\begin{table}[t!]
\caption{Supervised Segmentation analysis: Benchmarking BE and polyp datasets on standard segmentation architectures on 80-20 train-test split. \label{tab:1}} 
\centering 
\begin{tabular}{m{10em} m{5em} m{5em} | m{5em} m{5em}} %  
\toprule
\multirow{2}{*}{
\parbox[c]{.2\linewidth}{\centering Models}}
  & \multicolumn{2}{c}{BE} &
\multicolumn{2}{c}{Polyp} \\ 
\cmidrule{2-3} \cmidrule{4-5}
 &  \centering{\centering IoU} & \centering{\centering Dice} & \centering{\centering IoU} & {\centering Dice}  \\ 
\midrule
U-Net\cite{Olaf:MICCAI2015} &  \centering 0.738  & \centering0.846  &  \centering0.718 & 0.755 \\
UNet++\cite{zhou2018unet++} & \centering 0.742 & \centering 0.859 & \centering 0.721 & 0.778 \\
ResUNet\cite{diakogiannis2020resunet} & \centering 0.789 & \centering 0.908 & \centering 0.734 & 0.794 \\
EfficientUNet\cite{9150621} & \centering \textbf{0.841} & \centering \textbf{0.931} & \centering \textbf{0.803} & \textbf{0.896} \\
EfficientUNet++ & \centering 0.828 & \centering 0.927 & \centering 0.767 & 0.865 \\
DeepLabv3+ \cite{10.1007/978-3-030-01234-2_49}& \centering 0.802 & \centering 0.893 & \centering 0.789 & 0.869 \\
\bottomrule
\end{tabular}
\end{table}
%%%%%%%%%
\subsection{Benchmark comparison and ablation study}
% Supplementary material: Decoder, Sobel operator, Loss functions, Computational complexity
Result of supervised segmentation of BE and {polyps} in our GI dataset {are presented in Table~\ref{tab:1}}. 
{Frequently used supervised learning methods have been evaluated with respect to IoU and Dice to establish a baseline.} It can be observed that the UNet with {an} EfficientNet B4 backbone~\cite{9150621} outperformed other baseline methods including DeepLabv3+~\cite{10.1007/978-3-030-01234-2_49} by nearly 4\% and 2\% on both IoU and Dice for BE and {polyps}, respectively.~\textbf{Supplementary material Table 1-4} show that our choice of {the} joint loss function, {the} Sobel edge operator, {the} decoder architecture and {the} computational complexity yield in improved performance compared to {alternatives that are often used}.
%
%We first performed supervised segmentation analysis on both BE and polyps datasets using SOTA segmentation algorithms including UNet\cite{Olaf:MICCAI2015}, UNet++\cite{DBLP:journals/corr/abs-1807-10165}, ResUNet\cite{DBLP:journals/corr/abs-1904-00592}, EfficientUNet\cite{9150621}, EfficientUNet++, DeepLabv3+\cite{DBLP:journals/corr/abs-1802-02611}. Table 1 shows that EfficientUNet architecture which is a UNet framework with EfficientNet B4 backbone outputs the best performance in supervised semantic segmentation experiments, IoU=0.841 and IoU=0.803 for BE and polyps datasetets, respectively. This results in choosing EfficientUNet framework as segmentation network in our proposed EndoUDA pipeline for benchmark analysis with other SOTA UDA approaches.
%

\begin{table}[t!]
\caption{Empirical results of EndoUDA along with SOTA UDA methods with mean $\mu$ and standard deviation $\sigma$ are provided. All comparisons are provided for source only trained model and tested on target data. \textit{Paired t-test} between EndoUDA and each SOTA methods are shown in Supplementary Table 5.~\label{tab:2}}
\centering 
\begin{tabular}{m{10em} c c c c | c c c c} %  
\toprule
\multirow{2}{*}{
\parbox[c]{.2\linewidth}{\centering Models}}
  & \multicolumn{4}{c|}{BE ($\mu \pm \sigma$)}  &
\multicolumn{4}{c}{Polyp ($\mu \pm \sigma$)} \\ 
\cmidrule{2-5} \cmidrule{6-9}
 &  {IoU} & {Dice} & {Precision} & {Recall} & {IoU} & {Dice} & {Precision} & {Recall} \\ 
\midrule
Naive U-Net~\cite{9150621} &  \begin{tabular}[t]{@{}l@{}}0.626\\ \tiny${\pm{0.02}}$\end{tabular}  & \begin{tabular}[t]{@{}l@{}}0.718\\ \tiny${\pm{0.015}}$\end{tabular}  & \begin{tabular}[t]{@{}l@{}}0.654\\ \tiny${\pm{0.018}}$\end{tabular}   & \begin{tabular}[t]{@{}l@{}}0.667\\ \tiny${\pm{0.021}}$\end{tabular}  & \begin{tabular}[t]{@{}l@{}}0.492\\ \tiny${\pm{0.018}}$\end{tabular}  & \begin{tabular}[t]{@{}l@{}}0.564\\ \tiny${\pm{0.012}}$\end{tabular}  & \begin{tabular}[t]{@{}l@{}}0.487\\ \tiny${\pm{0.011}}$\end{tabular}   & \begin{tabular}[t]{@{}l@{}}0.496\\ \tiny${\pm{0.011}}$\end{tabular} \\
AdaptSegnet~\cite{Tsai} & \begin{tabular}[t]{@{}l@{}}0.658\\ \tiny${\pm{0.019}}$\end{tabular}  & \begin{tabular}[t]{@{}l@{}}0.749\\ \tiny${\pm{0.016}}$\end{tabular} & \begin{tabular}[t]{@{}l@{}}0.721\\ \tiny${\pm{0.012}}$\end{tabular}  & \begin{tabular}[t]{@{}l@{}}0.698\\ \tiny${\pm{0.022}}$\end{tabular} & \begin{tabular}[t]{@{}l@{}}0.519\\ \tiny${\pm{0.017}}$\end{tabular} & \begin{tabular}[t]{@{}l@{}}0.577\\ \tiny${\pm{0.015}}$\end{tabular} & \begin{tabular}[t]{@{}l@{}}0.601\\ \tiny${\pm{0.016}}$\end{tabular} & \begin{tabular}[t]{@{}l@{}}0.583\\ \tiny${\pm{0.012}}$\end{tabular}\\
ADVENT~\cite{Vu} & \begin{tabular}[t]{@{}l@{}}0.667\\ \tiny${\pm{0.012}}$\end{tabular} & \begin{tabular}[t]{@{}l@{}}0.768\\ \tiny${\pm{0.014}}$\end{tabular} & \begin{tabular}[t]{@{}l@{}}0.739\\ \tiny${\pm{0.015}}$\end{tabular}  & \begin{tabular}[t]{@{}l@{}}0.706\\ \tiny${\pm{0.012}}$\end{tabular} & \begin{tabular}[t]{@{}l@{}}0.524\\ \tiny${\pm{0.011}}$\end{tabular} & \begin{tabular}[t]{@{}l@{}}0.591\\ \tiny${\pm{0.019}}$\end{tabular} & \begin{tabular}[t]{@{}l@{}}0.662\\ \tiny${\pm{0.021}}$\end{tabular} & \begin{tabular}[t]{@{}l@{}}0.638\\ \tiny${\pm{0.014}}$\end{tabular}\\
CellSegUDA~\cite{pmlr-v121-haq20a} & \begin{tabular}[t]{@{}l@{}}0.673\\ \tiny${\pm{0.02}}$\end{tabular} &  \begin{tabular}[t]{@{}l@{}}0.771\\ \tiny${\pm{0.016}}$\end{tabular} & \begin{tabular}[t]{@{}l@{}}0.733\\ \tiny${\pm{0.018}}$\end{tabular} & \begin{tabular}[t]{@{}l@{}}0.691\\ \tiny${\pm{0.013}}$\end{tabular} & \begin{tabular}[t]{@{}l@{}}0.533\\ \tiny${\pm{0.012}}$\end{tabular} & \begin{tabular}[t]{@{}l@{}}0.629\\ \tiny${\pm{0.021}}$\end{tabular} & \begin{tabular}[t]{@{}l@{}}0.688\\ \tiny${\pm{0.018}}$\end{tabular} & \begin{tabular}[t]{@{}l@{}}0.641\\ \tiny${\pm{0.017}}$\end{tabular}\\
GLSS~\cite{10.1007/978-3-030-58539-6_25} & \begin{tabular}[t]{@{}l@{}}0.704\\ \tiny${\pm{0.019}}$\end{tabular} & \begin{tabular}[t]{@{}l@{}}0.815\\ \tiny${\pm{0.022}}$\end{tabular} & \begin{tabular}[t]{@{}l@{}}\textbf{0.846}\\ \tiny${\pm{0.019}}$\end{tabular} & \begin{tabular}[t]{@{}l@{}}0.735\\ \tiny${\pm{0.02}}$\end{tabular} & \begin{tabular}[t]{@{}l@{}}0.558\\ \tiny${\pm{0.017}}$\end{tabular} & \begin{tabular}[t]{@{}l@{}}0.621\\ \tiny${\pm{0.016}}$\end{tabular} & \begin{tabular}[t]{@{}l@{}}0.675\\ \tiny${\pm{0.019}}$\end{tabular} & \begin{tabular}[t]{@{}l@{}}0.649\\ \tiny${\pm{0.012}}$\end{tabular} \\
EndoUDA & \begin{tabular}[t]{@{}l@{}}\textbf{0.733}\\ \tiny${\pm{0.014}}$\end{tabular} &  \begin{tabular}[t]{@{}l@{}}\textbf{0.854}\\ \tiny${\pm{0.021}}$\end{tabular} & \begin{tabular}[t]{@{}l@{}}0.832\\ \tiny${\pm{0.017}}$\end{tabular} & \begin{tabular}[t]{@{}l@{}}\textbf{0.784}\\ \tiny${\pm{0.019}}$\end{tabular} & \begin{tabular}[t]{@{}l@{}}\textbf{0.605}\\ \tiny${\pm{0.014}}$\end{tabular} & \begin{tabular}[t]{@{}l@{}}\textbf{0.693}\\ \tiny${\pm{0.02}}$\end{tabular} & \begin{tabular}[t]{@{}l@{}}\textbf{0.722}\\ \tiny${\pm{0.017}}$\end{tabular} & \begin{tabular}[t]{@{}l@{}}\textbf{0.704}\\ \tiny${\pm{0.018}}$\end{tabular}\\
\bottomrule
\end{tabular}
\end{table}
%\subsection{Qualitative evaluation}
%%%%%%%%%
\begin{figure}[t!] % ALWAYS TOP FOR FIGURES AND TABLE IN ALL PAPERS! THAT's the rule!
\centering
\includegraphics[width=0.99\textwidth]{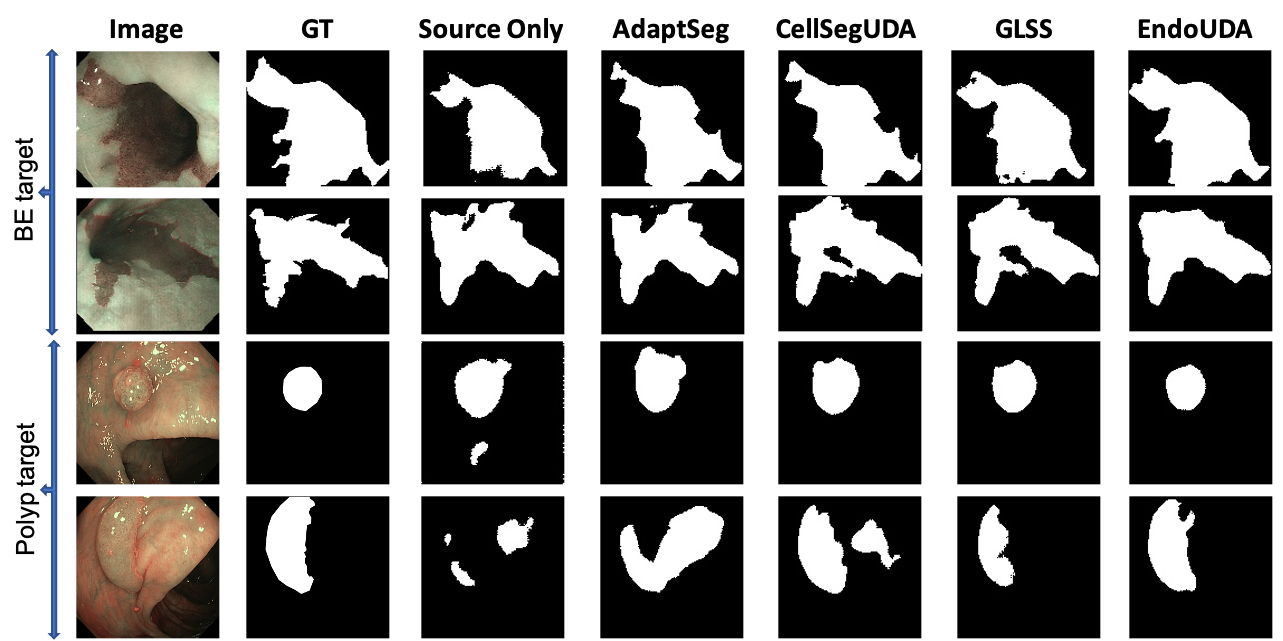}
\caption{Qualitative study: Baseline UDA experiments on BE and polyp images.~\label{fig:3}}
\end{figure}
%%%%%%%%%
\subsection{Comparison with SOTA method}
We compared the proposed EndoUDA approach with {four} SOTA UDA methods: AdaptSegnet\cite{Tsai}, ADVENT\cite{Vu}, CellSegUDA\cite{pmlr-v121-haq20a}, GLSS\cite{10.1007/978-3-030-58539-6_25}, along with the naive supervised baseline model. The quantitative results are reported in Table~\ref{fig:2} for four different metrics including intersection-over-union (IoU), Dice, precision and recall metrics. It can be observed that our proposed framework EndoUDA outperformed other SOTA UDA methods and naive U-Net method. An improvement of nearly 4\% on BE and 7\% on polyp dataset for {with respect to the Dice} metric compared to the recent SOTA GLSS method. Similar improvements can be seen for other metrics for both GI datasets and also much higher improvements compared to other SOTA UDA methods. {Table 5 in the supplementary material  shows the statistical significance of the improvement provided by our method when} compared to SOTA methods presented in Table~\ref{tab:2} with p-values $<$0.01 for almost all metrics and SOTA approaches for both BE and polyp segmentation. 

It can be seen in Fig.~\ref{fig:3} that the predicted masks using proposed EndoUDA model for BE and polyp images are very close to the ground truth annotations compared to those predicted by other SOTA UDA methods.

%It can be observed that our proposed framework outperformed the other baseline models p most of the metrics, improving IoU by 3\% and 4\% on both BE and polyps datasets, respectively. It can be seen in Fig. 3 that the predicted masks using proposed EndoUDA model for BE and polyps images are more similar to the ground truth images in comparison to those predicted by other UDA and (source only) models.
%
% \subsubsection{Ablation Study.}
% We conducted several ablation studies on EndoUDA using both BE and polyps datasets to examine the effect of various implementation of choices that we designed in our approach. The supplementary material shows the quantitative results of ablation studies that include the effect of the proposed normalized cross correlation loss ($\euscr{L}_{NCC}$); the effect of Sobel edge operator during training and inference time; the effect of coupling decoder of EfficientUnet into the decoder of VAE and comparison for computational complexity of the study. 
%
%%%%%%%%
%%%%%%%%%

\subsection{Effect of out-of-distribution data in supervised training}
We further investigated the performance of EndoUDA compared to the supervised baseline U-Net model (with EfficientNet backbone) by training the model with 10\%, 25\% and 50\% of target data samples. It can be observed {(see Table~\ref{tab:3}} that the EndoUDA performs consistently better than the baseline approach for {a} different proportion of target data inclusion in the training. For example, on 10\% and 25\% of target data mixed in the training sample, EndoUDA still {provides} nearly {a} 7-10\% improvement on IoU and 8-13\% {with respect to} Dice scores over the baseline method.
%As domain adaptation plays a key role in minimizing the domain shift problem, Table 3 shows the efficacy of our model that achieves greater performance than the baseline model when the same proportion of target annotations are applied. 
% As a comparison, we also trained the baseline EfficientUNet model with same proportion that is used for EndoUDA and also with fully supervised manner in that all target annotations are available. 
\begin{table}[t!]
\caption{IoU and dice comparisons for out-of-distribution target samples (NBI) in the source (WLI) GI endoscopy datasets~\label{tab:3}} 
\centering 
\begin{tabular}{l | c | c | c | c} %  
\toprule
Method & BE$_{IoU}$ & Polyp$_{IoU}$ & BE$_{dice}$ & Polyp$_{dice}$\\ 
\midrule
U-Net (source-only) & 0.626 & 0.492 & 0.718 & 0.564\\
EndoUDA & 0.733 & 0.605 & 0.854 & 0.693 \\
U-Net (source 100\% + target 10\%) & 0.646 & 0.536 & 0.738 & 0.588 \\
EndoUDA (source 100\% + target 10\%)  & 0.744 & 0.613 & 0.866 & 0.706 \\
U-Net (source 100\% + target 25\%) & 0.691 & 0.579 & 0.761 & 0.641 \\
EndoUDA (source 100\% + target 25\%) & 0.756 & 0.624 & 0.875 & 0.721 \\
U-Net (source 100\% + target 50\%) & 0.738 & 0.631 & 0.827 & 0.734 \\
EndoUDA (source 100\% + target 50\%) & 0.771 & 0.658 & 0.889 & 0.739 \\
U-Net (target-trained) & 0.841 & 0.802 & 0.931 & 0.896 \\
\bottomrule
\end{tabular}
\end{table}
%%%%%%%%%
\section{Conclusion}
%%%%%%%%
We have presented a novel target-independent unsupervised domain adaptation VAE-based method that comprises of a shared encoder for both reconstruction and segmentation, and a joint loss function for improved unseen target domain generalization. We have validated our approach on two GI endoscopy datasets that often require additional modalities for screening and intervention. The proposed modality agnostic method can be used in clinical GI endoscopy without imposing {any constraints on what imaging modality is being used.} Our qualitative and quantitative results demonstrate the effectiveness and promise of the proposed method compared to both the naive baseline supervised model and state-of-the-art UDA segmentation methods.

\subsubsection*{Acknowledgement}
NC is supported by the Emerson Collective Cancer Research Fund, SG is funded by Boston Scientific, SA and JR are supported by National Institute for Health Research (NIHR) Oxford Biomedical Research Centre (BRC). The views expressed are those of the authors and not necessarily those of the NHS, the NIHR or the Department of Health. The authors declare no competing interests.

% In this work, we propose an unsupervised domain adaptation framework for segmentation of BE and polyp. This approach provides a generalized prediction of segmentation masks of unlabelled endoscopy images in cross modalities. The method does not depend on the existence of target labels and hence provides accurate and robust results in different imaging conditions. The experimental results demonstrate the effectiveness of the proposed UDA model by achieving better results than the state-of-the-art unsupervised domain adaptation methods. 
\bibliographystyle{splncs04}
\bibliography{ref}
\end{document}